\PassOptionsToPackage{dvipsnames}{xcolor}
\documentclass[runningheads]{llncs}
\usepackage{graphicx}
\usepackage{latexsym}
\usepackage{xcolor}
\usepackage{amsfonts}
\usepackage{amsmath}
\usepackage{enumitem}
\usepackage[usestackEOL]{stackengine}
\usepackage{xspace}
 \usepackage{booktabs}
 \usepackage{comment}
\usepackage{hhline}
\usepackage{multirow}

\usepackage{tikz}
\usepackage{pgfplots}
\usepgfplotslibrary{groupplots}
\usepackage{graphicx}
\usepackage{verbatim}
\usepackage{braket}
\usepackage{caption}
\usepackage{subcaption}
\usepackage{bm}
\usepackage{tabularx}
\usepackage{float}
\makeatletter
\newcommand*\bigcdot{\mathpalette\bigcdot@{.7}}
\newcommand*\bigcdot@[2]{\mathbin{\vcenter{\hbox{\scalebox{#2}{$\m@th#1\bullet$}}}}}
\makeatother

\definecolor{ballblue}{rgb}{0.13, 0.67, 0.8}
\definecolor{carnelian}{rgb}{0.7, 0.11, 0.11}
\usetikzlibrary{shapes.geometric}

\definecolor{col1}{RGB}{254, 69, 107}
\definecolor{col2}{RGB}{235, 72, 120}
\definecolor{col3}{RGB}{215, 75, 134}
\definecolor{col4}{RGB}{196, 78, 147}
\definecolor{col5}{RGB}{177, 81, 171}
\definecolor{col6}{RGB}{158, 85, 175}
\definecolor{col7}{RGB}{138, 88, 188}
\definecolor{col8}{RGB}{119, 91, 202}
\definecolor{col9}{RGB}{100, 94, 215}
\definecolor{col10}{RGB}{80, 97, 229}
\definecolor{col11}{RGB}{63, 100, 241}

\usepackage[htt]{hyphenat}

\def\framework{FedeRank\xspace}
\def\ml{\textit{MovieLens 1M}\xspace}
\def\am{\textit{Amazon Digital Music}\xspace}
\def\lb{\textit{LibraryThing}\xspace}
\def\library{\textit{LibraryThing}\xspace}

\def\adm{\textit{Amazon Digital Music}\xspace}
\newif\ifnotes
\notestrue

\ifnotes
\newcommand{\antonio}[1]{\textcolor{blue}{#1}}
\newcommand{\yashar}[1]{\textcolor{green!55!blue}{{\bf [Yashar: }{\em #1}{\bf ]}}}
\newcommand{\tommaso}[1]{\textcolor{magenta}{{\bf [Tommaso: }{\em #1}{\bf ]}}}
\newcommand{\lucio}[1]{\textcolor{violet}{{\bf [Lucio: }{\em #1}{\bf ]}}}
\newcommand{\walter}[1]{\textcolor{red}{{\bf [Walter: }{\em #1}{\bf ]}}}
\newcommand{\claudio}[1]{\textcolor{orange}{{\bf [Claudio: }{\em #1}{\bf ]}}}
\def\wanote#1{\todo[size=\scriptsize,backgroundcolor=white,linecolor=red,bordercolor=red]{\textit{WA}: #1}} 
\else
\newcommand{\antonio}[1]{}
\newcommand{\yashar}[1]{}
\newcommand{\tommaso}[1]{}
\newcommand{\lucio}[1]{}
\newcommand{\walter}[1]{}
\newcommand{\claudio}[1]{}
\def\wanote#1{\todo[size=\scriptsize,backgroundcolor=white,linecolor=red,bordercolor=red]{}} 
\fi

\newcommand{\dquotes}[1]{``#1''}

\usetikzlibrary{arrows,chains,matrix,positioning,scopes}
\makeatletter
\tikzset{join/.code=\tikzset{after node path={%
\ifx\tikzchainprevious\pgfutil@empty\else(\tikzchainprevious)%
edge[every join]#1(\tikzchaincurrent)\fi}}}
\makeatother
\tikzset{>=stealth',every on chain/.append style={join},
         every join/.style={->}}
\tikzstyle{labeled}=[execute at begin node=$\scriptstyle,
   execute at end node=$]

\pgfplotsset{compat=1.16}

\begin{document}
\title{\framework: User Controlled Feedback\\with Federated Recommender Systems}
%
%

\author{Vito Walter Anelli \and
Yashar Deldjoo \and Tommaso Di Noia \and Antonio Ferrara\thanks{Corresponding author} \and Fedelucio Narducci}
\authorrunning{V. W. Anelli et al.}
%
\institute{Politecnico di Bari, Bari, Italy\\
\email{\{firstname.lastname\}@poliba.it}}

\maketitle              
\begin{abstract}
Recommender systems have shown to be a successful representative of how data availability can ease our everyday digital life.
However, data privacy is one of the most prominent concerns in the digital era. After several data breaches and privacy scandals, the users are now worried about sharing their data. In the last decade, Federated Learning has emerged as a new privacy-preserving distributed machine learning paradigm. 
It works by processing data on the user device without collecting data in a central repository.
We present \mbox{FedeRank} (\url{https://split.to/federank}), a federated recommendation algorithm. The system learns a personal factorization model onto every device. The training of the model is a synchronous process between the central server and the federated clients.
FedeRank takes care of computing recommendations in a distributed fashion and allows users to control the portion of data they want to share.
By comparing with state-of-the-art algorithms, extensive experiments show the effectiveness of FedeRank in terms of recommendation accuracy, even with a small portion of shared user data.
Further analysis of the recommendation lists' diversity and novelty guarantees the suitability of the algorithm in real production environments.


\end{abstract}
%
%

\section{Introduction}
\label{sec:introduction}
Recommender Systems (RSs) are well-known information-filtering systems widely adopted for mitigating the information-overload problem. Indeed, the broad amount of items and services has caused a cognitive impairment that takes the name of over-choice, or choice overload. 
RSs have proved to be very useful in making possible personalized access to these catalogs of items. 
These systems are generally hosted on centralized servers and train their models by exploiting massive proprietary and sensitive data. 
However, public awareness related to data collection was spurred and increased. 
In recent years, an increasing number of countries have introduced regulations to protect user privacy and data security. Representative examples are the GDPR in the European Union~\cite{EUdataregulations2018}, the CCPA in California~\cite{CCPA}, and the Cybersecurity Law in China~\cite{chinacybersecurity}.
Such policies prohibit free data circulation and force personal data to remain isolated and fragmented. 

In this context, Google has recently proposed Federated Learning (FL) as a privacy-by-design technique which tackles data isolation while meeting the need for big data~\cite{DBLP:journals/corr/KonecnyMRR16,DBLP:conf/aistats/McMahanMRHA17}.
FL trains a global machine-learning model by leveraging both users' data and personal devices' computing capabilities.
Unlike previous approaches, it keeps data on the devices (e.g., smartphones, tablets, etc.) without sharing it with a central server. 
Today, FL is considered the best candidate to face the data privacy, control and property challenges posed by the data regulations.

Among the recommendation paradigms proposed in the literature, Collaborative Filtering (CF) demonstrated a very high accuracy~\cite{DBLP:journals/taslp/McFeeBL12,DBLP:conf/bigdataconf/YuanSKLAL16}.
The strength of CF recommendation algorithms is that users who expressed similar tastes in the past tend to agree in the future as well.
One of the most prominent CF approaches is the Latent Factor Model (LFM)~\cite{DBLP:reference/sp/KorenB15}. LFMs uncover users and items latent representation, whose linear interaction can explain observed feedback.

In this paper, we introduce FedeRank, a novel factorization model that embraces the FL paradigm.
A disruptive effect of employing FedeRank is that users participating in the federation process can decide if and how they are willing to disclose their private sensitive preferences. 
Indeed, FedeRank mainly leverages non-sensitive information (e.g., items the user has not experienced). 
Here, we show that even only a small amount of sensitive information (i.e., items the user has experienced) lets FedeRank reach a competitive accuracy.
How incomplete data impacts the performance of the system is an entirely unexplored field.
Analogously, it is still to establish the minimum amount of data necessary to build an accurate recommendation system~\cite{yang2019federated}. 
At the same time, preserving privacy at the cost of a worse tailored recommendation may frustrate users and reduce the \dquotes{acceptance of the recommender system}~\cite{DBLP:conf/kdd/MuhammadWOTSHGL20}.
In this work, instead of focusing on how to protect personal information from privacy breaches (that is explored in other active research fields), we investigate how to guarantee the users the control and property of their data as determined by regulations.
The work's contributions are manifold due to the number of open challenges that still exist with the FL paradigm.
To summarize, our contributions in this paper include:
\begin{itemize}
    \item the development of the first, to the best of our knowledge, federated pair-wise recommendation system;
    \item an analysis of the impact of client-side computation amount;
    \item an investigation on the existing relation between incomplete data and recommendation accuracy, and an analysis of the algorithmic bias on the final recommendation lists, based on the data deprivation amount.
\end{itemize}
To this extent, we have carried out extensive experiments on three real-world datasets (\am, \lb, and \ml) by considering two evaluation criteria: (a) the accuracy of recommendations measured by exploiting precision and recall, (b) beyond-accuracy measures to evaluate the novelty, and the diversity of recommendation lists.
The experimental evaluation shows that \framework provides high-quality recommendations, even though it leaves users in control of their data.

\section{Related Work}
\label{sec:related}
In the last decades, academia and industry have proposed several competitive recommendation algorithms.
Among the Collaborative Filtering algorithms, the most representative examples are undoubtedly Nearest Neighbors systems, Latent Factor Models, and Neural Network-based recommendation systems.
The Nearest Neighbors scheme has shown its competitiveness for decades.
After them, factorization-based recommendation emerged thanks to the disruptive idea of Matrix Factorization (MF). Nevertheless, several generalized/specialized variants have been proposed, such as FM~\cite{DBLP:conf/icdm/Rendle10}, SVD++~\cite{DBLP:conf/kdd/Koren08}, PITF~\cite{DBLP:conf/wsdm/RendleS10}, FPMC~\cite{DBLP:conf/www/RendleFS10}.
Unfortunately, rating-prediction-oriented optimization, like SVD, has shown its limits in the recommendation research~\cite{DBLP:conf/chi/McNeeRK06}. Consequently, a new class of \textit{Learning to Rank} algorithms has been developed in the last decade, mainly ranging from point-wise~\cite{DBLP:conf/recsys/KorenS11} to pair-wise \cite{DBLP:conf/uai/RendleFGS09}, through list-wise \cite{DBLP:conf/recsys/ShiLH10} approaches. Among pair-wise methods, BPR~\cite{DBLP:conf/uai/RendleFGS09} is one of the most adopted, thanks to its outstanding capabilities to correctly rank with an acceptable computational complexity.
Finally, in the last years, methods exploiting the architectures of deep neural networks have established themselves in search and recommendation research.

To make RSs work properly (easing the user decision-making process and boosting the business), they need to collect user information related to attributes, demands, and preferences~\cite{DBLP:series/ccn/JeckmansBEHLT13},
jeopardizing the user's privacy.
In this scenario --- and, more generally, in any scenario with a system learning from sensitive data --- FL was introduced for meeting privacy shortcomings by horizontally distributing the model’s training over user devices~\cite{DBLP:conf/aistats/McMahanMRHA17}. 
Beyond privacy, FL has posed several other challenges and opened new research directions~\cite{49232}.
In the last years, it has extended to a more comprehensive idea of privacy-preserving decentralized collaborative ML approaches~\cite{DBLP:journals/tist/YangLCT19}, ranging from horizontal federations, where different devices (and local datasets) share the same feature space, to vertical federations, where devices share training samples that differ in feature space.

Some researchers focused the attention on the decentralized and distributed matrix-factorization approaches~\cite{DBLP:conf/recsys/DuriakovaTSHPSG19,DBLP:journals/tnn/FierimonteSUP17}.
However, in this work, we focus on federated learning principles theoretically and practically different from classical distributed approaches.
Indeed, FL assumes the presence of a coordinating server and the use of private and self-produced data on each node.
In general, distributed approaches do not guarantee these assumptions.
Ammad-ud-din\textit{ et al.}~\cite{DBLP:journals/corr/abs-1901-09888} propose a federated implementation of collaborative filtering, whose security limits are analyzed in~\cite{9162459}, which uses the SVD-MF method for implicit feedback~\cite{DBLP:conf/icdm/HuKV08}. Here, the training is a mixture of Alternating Least Squares (ALS) and Stochastic Gradient Descent (SGD) for preserving users' privacy.
Nevertheless, incomprehensibly, almost no work addressed top-N recommendation exploiting the \dquotes{Learning to rank} paradigm.
In this sense, one rare example is the work by Kharitonov\textit{ et al.}~\cite{kharitonov2019federated}, who recently proposed to combine evolution strategy optimization with a privatization procedure based on differential privacy.
The previous work focuses neither on search or recommendation. Perhaps, like ours, it can be classified as a federated learning-to-rank algorithm.  
Finally,  Yang\textit{ et al.}~\cite{yang2019federated} identified some recent FL challenges and open research directions.

\section{Approach}
\label{sec:background}
In this section, we introduce the fundamental concepts regarding the Collaborative Filtering recommendation using a Federated Learning scheme. 
Along with the problem definition, the notation we adopt is presented.

The recommendation problem over a set of users $\mathcal{U}$ and a set of items $\mathcal{I}$ is defined as the activity of finding for each user $u \in \mathcal{U}$ an item $i \in \mathcal{I}$ that maximizes a utility function $g : \mathcal{U} \times \mathcal{I} \rightarrow \mathbb{R}$~\cite{DBLP:reference/sp/NingDK15}.
Let $\mathbf{X} \in\mathbb{R}^{|\mathcal{U}| \times |\mathcal{I}|}$ be the user-item matrix containing for each element $x_{ui}$ an implicit feedback (e.g., purchases, visits, clicks, views, check-ins) of user $u \in \mathcal{U}$ for item $i \in \mathcal{I}$.
Therefore, $\mathbf{X}$ only contains binary values, $x_{ui} = 1$ and $x_{ui} = 0$ denoting whether user $u$ has consumed or not item $i$, respectively. 

The recommendation model is based on Factorization approach, originally introduced by Matrix Factorization~\cite{DBLP:journals/computer/KorenBV09}, that became popular in the last decade thanks to its state-of-the-art recommendation accuracy~\cite{DBLP:journals/corr/BokdeGM15a}. 
This technique aims to build a model $\Theta$ in which each user $u$ and each item $i$ is represented by the embedding vectors $\mathbf{p}_u$ and $\mathbf{q}_i$, respectively, in the shared latent space $\mathbb{R}^F$. Let assume $\mathbf{X}$ can be factorized such that the dot product between $\mathbf{p}_u$ and $\mathbf{q}_i$ can explain any observed user-item interaction $x_{ui}$, and any non-observed interaction can be estimated as
$\hat{x}_{ui}(\Theta) = b_i(\Theta) + \mathbf{p}_u^T(\Theta)\cdot \mathbf{q}_i(\Theta) $
where $b_i$ is a term denoting the bias of the item $i$.

Among pair-wise approaches for learning-to-rank the items of a
catalog, Bayesian Personalized Ranking~\cite{DBLP:conf/uai/RendleFGS09} is the most broadly adopted, thanks to its capabilities to correctly rank with \textit{acceptable} computational complexity.
Given a training set defined by $\mathcal{K}=\{(u,i,j) \;|\; x_{ui} = 1 \land x_{uj} = 0 \}$, 
 BPR minimizes the ranking loss by exploiting the criterion $\underset{\Theta}{\max}\ G(\Theta)$, with $G(\Theta) = \sum_{(u,i,j) \in \mathcal{K}} \ln \ \sigma (\hat{x}_{uij}(\Theta)) - \lambda  \lVert \Theta \rVert^2$,
where $\hat{x}_{uij}(\Theta) = \hat{x}_{ui}(\Theta) - \hat{x}_{uj}(\Theta)$ is a real value modeling the relation between user $u$, item $i$ and item $j$, $\sigma(\cdot)$ is the sigmoid function, and $\lambda$ is a model-specific regularization parameter to prevent overfitting.
Pair-wise optimization applies to a wide range of recommendation models, including factorization. 
Hereafter, we denote the  model $\Theta = \Braket{\mathbf{P}, \mathbf{Q}, \mathbf{b}}$, where $\mathbf{P} \in \mathbb{R}^{|\mathcal{U}|\times F}$ is a matrix whose $u$-th row corresponds to the vector $\mathbf{p}_u$, and $\mathbf{Q} \in \mathbb{R}^{|\mathcal{I}|\times F}$ is a matrix in which the $i$-th row corresponds to the vector $\mathbf{q}_i$. Finally, $\mathbf{b} \in \mathbb{R}^{|\mathcal{I}|}$ is a vector whose $i$-th element corresponds to the value $b_i$.

\subsection{\framework}
\framework redesigns the original factorization approach for a federated setting.
Indeed, the initial factorization model and its variants use a single, centralized model, which does not guarantee users to control their data. \framework splits the pair-wise learning model $\Theta$ among a central server $S$ and a federation of users $\mathcal{U}$.
Federated learning aims to optimize a global loss function by using data distributed among a federation of users' devices. The rationale is that the server no longer collects private users' data. Rather, it aggregates the results of some steps of local optimizations performed by clients, preserving privacy, ownership, and locality of users' data~\cite{DBLP:journals/corr/abs-1902-01046}.
Formally, let $\Theta$ be the machine learning model parameters, and $G(\Theta)$ be a loss function to minimize. 
In Federated learning, the users $\mathcal{U}$ of a federation collaborate to minimize $G$ (under the coordination of a central server $S$) without sharing or exchanging their raw data.
From an algorithmic point of view, $S$ shares $\Theta$ with the federation of devices. Then, the optimization problem of minimizing $G$ is locally solved.
Since each user participates to the federation with her personal data and with her personal client device, we will interchangeably use the terms \dquotes{client}, \dquotes{user}, and \dquotes{device}.

To set up the framework, we consider the central server $S$ holding a model $\Theta_S = \Braket{\mathbf{Q}, \mathbf{b}}$, where $\mathbf{Q} \in \mathbb{R}^{|\mathcal{I}|\times F}$ is a matrix in which $i$-th row represents the embedding $\mathbf{q}_i$ for item $i$ in the catalog, while the element $b_i$ of $\mathbf{b} \in \mathbb{R}^{|\mathcal{I}|}$ is the bias of item $i$. That is, the information on $S$ only characterizes the items of the catalog.
On the other hand, each user $u \in \mathcal{U}$ holds a local model $\Theta_u = \Braket{\mathbf{p}_u}$, where $\mathbf{p}_u \in \mathbb{R}^{F}$ corresponds to the representation of user $u$ in the latent space of dimensionality $F$. Each user holds a private interaction dataset $\mathbf{x}_u \in \mathbb{R}^{|\mathcal{I}|}$, which --- compared to a centralized recommender system --- corresponds to the $\mathbf{X}$'s $u$-th row.
The user $u$ leverages her private dataset $\mathbf{x}_u$ to build the local training set $\mathcal{K}_u=\{(u,i,j) \;|\; x_{ui}=1 \land x_{uj}=0 \}$. Finally, the overall number of interactions in the system can be obtained by exploiting the local datasets. Let us define it as $X^+ = \sum_{u \in \mathcal{U}} |\{x_{ui} | x_{ui} = 1\}|$.

The training procedure iterates for $E$ \textit{epochs}, in each of which \textit{rpe} \textit{rounds of communication} between the server and the devices are performed. A round of communication envisages a \textbf{Distribution to Devices} $\rightarrow$ \textbf{Federated Optimization} $\rightarrow$ \textbf{Transmission to Server} $\rightarrow$ \textbf{Global Aggregation} sequence.
The notation $\{\cdot\}_S^t$ denotes an object computed by the server $S$ at round $t$, while $\{\cdot\}_u^t$ indicates an object computed by a specific client $u$ at round $t$.

    \noindent (1) \textbf{Distribution to Devices.} Let $\{\mathcal{U}^-\}_S^t$ be a subset of $\mathcal{U}$ with cardinality $m$, containing $m$ clients $u \in \mathcal{U}$. The set $\{\mathcal{U}^-\}_S^t$ can be either defined by $S$, or the result of a request for availability sent by $S$ to clients in $\mathcal{U}$. Each client $u \in \{\mathcal{U}^-\}_S^t$ receives from $S$ the latest snapshot of $\{\Theta_S\}_S^{t-1}$.

    \noindent (2) \textbf{Federated Optimization.} Each user $u \in \{\mathcal{U}^-\}_S^t$ generates the set $\{\mathcal{K}_u^-\}_u^t$ containing $T$ random triples $(u, i, j)$ from $\mathcal{K}_u$. It is worth underlining that Rendle~\cite{DBLP:conf/uai/RendleFGS09} suggests, for a centralized scenario, to train the recommendation model by randomly choosing the training triples from $\mathcal{K}$, to avoid data is traversed item-wise or user-wise, since this may lead to slow convergence. Conversely, in a federated approach, we require to train the model user-wise. Indeed, the learning is separately performed on each device ($u$), that only knows the data in $\mathcal{K}_u$. Thanks to the user-wise traversing, \framework can decide who controls (the designer or the user) the number of triples $T$ in the training set $\{\mathcal{K}_u^-\}_u^t$, to tune the degree of local computation.
    With the local training set, the user $u$ can compute her contribution to the overall optimization of $\Theta_S$ with the following update:
    \begin{equation}
    \label{eq:update_thetas}
    \{\Delta\Theta_S\}_u^t = \{ \Delta \Braket{\mathbf{Q},\mathbf{b}} \}_u^t
    \mathrel{\mathop:}=
    \sum_{(u, i, j) \in 
    \{\mathcal{K}_u^-\}_u^t}
    \frac{\partial}{\partial \Theta_S}
    \ln \ \sigma (\hat{x}_{uij}(\{\Theta_S\}_S^{t-1}; \{\Theta_u\}_u^{t-1})),
    \end{equation}
    plus a regularization term. At the same time, the client $u$ updates its local model $\Theta_u$, and incorporates it in the current model by using:
    \begin{equation}
    \label{eq:update_client}
    \{\Delta\Theta_u\}_u^t = \{\Delta\Braket{\mathbf{p}_u}\}_u^t
    \mathrel{\mathop:}=
    \sum_{(u, i, j) \in 
    \{\mathcal{K}_u^-\}_u^t}
    \frac{\partial}{\partial \Theta_u}
    \ln \ \sigma (\hat{x}_{uij}(\{\Theta_S\}_S^{t-1}; \{\Theta_u\}_u^{t-1})),
    \end{equation}
    plus a regularization term. The partial derivatives in Eq.~\ref{eq:update_thetas} and \ref{eq:update_client} are straightforward, and can be easily computed by following the scheme proposed by Rendle \textit{et al.}~\cite{DBLP:conf/uai/RendleFGS09}.
    At the end of the federated computation, given a shared learning rate $\alpha$, each client can update its local model $\Theta_u$ --- containing the user profile --- by aggregating the computed update:
    \begin{equation}
    \{\Theta_u\}_u^t
    \mathrel{\mathop:}=
    \{\Theta_u\}_u^{t-1} + 
    \alpha
    \{\Delta\Theta_u\}_u^t.
    \end{equation}
    
    \noindent (3) \textbf{Transmission to Server.} In a purely distributed architecture, each user in $\mathcal{U}^-$ returns to $S$ the computed update. Here, instead of sending $\{ \Delta \Theta_S \}_u^t$, each user transmits a modified version $\{ \Delta \Theta_S^\Phi \}_u^t$.
    To introduce this aspect of \framework, let us define $\mathcal{F} = \{i, \, \forall (u, i, j) \in \{\mathcal{K}_u^-\}_u^t \}$,
    and a randomized object $\Phi = \Braket{\mathbf{Q}^\Phi, \mathbf{b}^\Phi}$, with $\mathbf{Q}^\Phi \in \mathbb{R}^{|\mathcal{I}|\times F}$, and $\mathbf{b}^\Phi \in \mathbb{R}^{|\mathcal{I}|}$.
    Each row $\mathbf{q}_i^\Phi$ of $\mathbf{Q}^\Phi$ and each element $b_i^\Phi$ of $\mathbf{b}^\Phi$ assume their value according to the probabilities:
    \begin{equation}
        \begin{gathered}
        P(\mathbf{q}_i^\Phi = \mathbf{1}, b_i^\Phi = 1 \mid i \in \mathcal{F}) = \pi, \quad
            P(\mathbf{q}_i^\Phi = \mathbf{0}, b_i^\Phi = 0 \mid i \in \mathcal{F}) = 1 - \pi, \\
                P(\mathbf{q}_i^\Phi = \mathbf{1}, b_i^\Phi = 1 \mid i \notin \mathcal{F}) = 1
    \end{gathered}
    \end{equation}
    Based on $\{ \mathbf{Q}^\Phi \}_u^t$ and $\{ \mathbf{b}^\Phi \}_u^t$, $\Delta \Theta_S^\Phi$ can be computed as it follows:
    \begin{equation}
    \label{eq:privacy}
        \{ \Delta \Theta_S^\Phi \}_u^t = \{ \Delta \Theta_S \}_u^t \odot \{ \Phi \}_u^t \mathrel{\mathop:}= \Braket{\{ \Delta \mathbf{Q} \}_u^t \odot \{ \mathbf{Q}^\Phi \}_u^t, \{ \Delta \mathbf{b} \}_u^t \odot \{ \mathbf{b}^\Phi \}_u^t},
    \end{equation}
    where the operator $\odot$ denotes the Hadamard product.
    This transformation is motivated by a possible privacy issue.
    The update $\Delta \mathbf{Q}$ computed in Eq.~\ref{eq:update_thetas} by user $u$ is a matrix whose rows are non-zero in correspondence of the items $i$ and $j$ of all the triples $(u, i, j) \in \mathcal{K}_u^-$~\cite{DBLP:conf/uai/RendleFGS09}. An analogous behavior can be observed for the elements of $\Delta \mathbf{b}$.
    Focusing on the non-zero elements, we observe that, for each triple $(u, i, j) \in \mathcal{K}_u^-$, the updates $\{\Delta \mathbf{q}_i\}_u^t$ and $\{\Delta \mathbf{q}_j\}_u^t$, as well as $\{\Delta b_i\}_u^t$ and $\{\Delta b_j\}_u^t$, show the same absolute value with opposite sign~\cite{DBLP:conf/uai/RendleFGS09}. In fact, this makes completely distinguishable for the server the consumed and the non-consumed items of user $u$, allowing $S$ to reconstruct $\mathcal{K}_u^-$, thus raising a privacy issue.
    Since our primary goal is to put users in control of their data, we leave users the possibility to choose a fraction $\pi$ of positive item updates to send. The remaining positive item updates (a fraction $1-\pi$) are masked by setting them to zero, by means of the transformation in Eq.~\ref{eq:privacy}.
    Conversely, the negative updates are always sent to $S$, since their corresponding rows are always multiplied by a $\mathbf{1}$ vector. Indeed, these updates are related to non-consumed items, which are indistinguishably negative or missing values, assumed to be \textit{non-sensitive} data.

    \noindent(4) \textbf{Global Aggregation.} Once $S$ has received $\{\Delta\Theta_S^\Phi\}_u^t$ from all clients $u \in \mathcal{U}^-$, it aggregates the received updates in $\mathbf{Q}$ and $\mathbf{b}$ to build the new global model, with $\alpha$ being the learning rate:
    \begin{equation}
    \{\Theta_S\}_S^t
    \mathrel{\mathop:}=
    \{\Theta_S\}_S^{t-1} + 
    \alpha \sum_{u \in \mathcal{U}^-}
    \{\Delta\Theta_S^\Phi\}_u^t.
    \end{equation}

Although Federated Learning was conceived as a privacy-by-design paradigm for distributed machine learning, it still does not provide formal privacy guarantees. Malicious actors might acquire different information. They may be able to analyze remote devices or communication flows in the network or infer users' private data by inspecting updates received on the server~\cite{49232}. 
Possible solutions include the encryption of data on local devices, the network traffic, or the adoption of secure multi-party computation~\cite{DBLP:conf/ccs/BonawitzIKMMPRS17}. 
Moreover, local differential privacy can guarantee that even if an adversary can inspect the communication between a user and the central server, she can learn only limited information~\cite{DBLP:journals/fttcs/DworkR14,DBLP:conf/ccs/ErlingssonPK14,appleprivacy}. 
To date, \framework explicitly provides the needed APIs to work, out of the box, with encryption communication libraries, thus providing state-of-the-art privacy guarantees.
We have chosen this solution since discussing privacy and security implications in FL is beyond our scope. In this way, the system designer can choose the privacy solution while disregarding the underlying machine learning model.
Moreover, \framework can also be easily adapted to guarantee local differential privacy. Indeed, it is not due to chance the choice of putting the user in control of $\Phi$. 
Suppose the $\Phi$ object also considers sending fake information. In that case, \framework becomes utterly compliant with the randomized response technique, which guarantees differential privacy~\cite{DBLP:conf/edbt/0009WH16}.

\section{Experiments}
\label{sec:experiments}
\vspace{-0.3em}
\noindent \textbf{Datasets.} 
We have investigated the performance of \framework considering three well-known datasets: \am~\cite{mcauley2015image}, \lb~\cite{zhao2015improving}, and \ml~\cite{harper2015movielens}. The former includes the users' satisfaction feedback for a catalog of music tracks available with Amazon Digital Music service. It contains 1,835 users and 41,488 tracks, with 75,932 ratings ranging from $1$ to $5$. \lb collects the users' ratings on a book catalog. It captures the interactions of 7,279 users on 37,232 books. It provides more than two million ratings with 749,401 unique ratings in a range from $1$ to $10$. The latter is \ml dataset, which collects users' ratings in the movie domain: it contains 1,000,209 ratings ranging from $1$ to $5$, 6,040 users, and 3,706 items. We have filtered out users with less than $20$ ratings (considering them as cold-users). Table \ref{tbl:dataset_char}
shows the characteristics of the resulting datasets adopted in the experiments.

{\setlength{\tabcolsep}{0.38em}
\begin{table}[t]
  \caption{Characteristics of the datasets used in the offline experiments: $\left| \mathcal{U} \right|$ is the number of users, $\left| \mathcal{I} \right|$ the number of items, $X^+$ the amount of positive feedback.} 
  \centering
  \label{tbl:dataset_char}
\scriptsize
  \begin{tabular}{ccccccc}
    \toprule
    \textbf{Dataset} & $\left| \mathcal{U} \right|$ & $\left| \mathcal{I} \right|$ & $X^+$ &$\frac{X^+}{\left| \mathcal{U} \right|}$ & $\frac{X^+}{\left| \mathcal{I} \right|}$ & $\frac{X^+}{\left| \mathcal{I} \right| \cdot \left| \mathcal{U} \right|} \%$\\
    \toprule
    \textbf{Amazon DM} & 1,835 & 41,488 & 75,932 & 41.38 & 1.83 & 0.000997\% \\
    \textbf{LibraryThing} & 7,279 & 37,232 & 749,401 & 102.95 & 20.13 &  0.002765\% \\
    \textbf{MovieLens 1M} & 6,040 & 3,706 & 1,000,209 & 165.60 & 269.89 & 0.044684\% \\
    \bottomrule 
  \end{tabular}%
\end{table}
}

\noindent \textbf{Baseline Algorithms.} 
We compared \framework with representative centralized algorithms to position its performance with respect to the state-of-the-art techniques: \textbf{VAE}~\cite{liang2018variational}, a non-linear probabilistic model taking advantage of Bayesian inference to estimate the model parameters;
\textbf{User-kNN} and \textbf{Item-kNN}~\cite{koren2010factor}, two neighbor-based CF algorithms, that exploit cosine similarity to compute similarity between users or items; \textbf{BPR-MF}~\cite{DBLP:conf/uai/RendleFGS09}, the centralized vanilla BPR-MF implementation; and \textbf{FCF}~\cite{DBLP:journals/corr/abs-1901-09888}, the only federated recommendation approach, to date, based on MF\footnote{Since no source code is available, we reimplemented it in the reader's interest.}. 
We have evaluated \framework considering $|\mathcal{U}^-|=1$. That is, in each round of communication we involve only a single client to avoid noisy results. 
We thereby guarantee the sequential training, needed to compare against centralized pir-wise techniques. We have investigated with two different \framework settings. In the \textbf{first setting}, we have set $T=1$, i.e., each client extracts solely one triple $(u, i, j)$ from its dataset when asked for training the model; with this special condition, we test whether \framework is effectively comparable to BPR. Moreover, to make the comparison completely fair, we extract triples as proposed by Rendle \textit{et al.}~\cite{DBLP:conf/uai/RendleFGS09}.
The \textbf{second setting} follows a real Federated scenario where the client local computation is not limited to a single triple. Specifically, the number $T$ of triples extracted by each client is set to $\frac{X^+}{|\mathcal{U}|}$.

\noindent \textbf{Reproducibility and Evaluation Metrics.} 
To train \framework, we have adopted a realistic temporal hold-out $80$-$20$ splitting for training set and test set \cite{DBLP:reference/sp/GunawardanaS15}.
We have further split the training set adopting a temporal hold-out strategy on a user basis to pick the last 20\% of interactions as a validation set. Hence, we have explored a grid in the range $\{0.005,\dots,0.5\}$. Then, to ensure a fair comparison, we have used the same learning rate to train \framework. We have set up the remaining parameters as follows: the user- and positive item-regularization parameter is set to $1/20$ of the learning rate; conversely, the negative item-regularization parameter is set to $1/200$ of the learning rate as suggested by Anelli \textit{et al.} \cite{DBLP:conf/recsys/AnelliNSPR19}. Moreover, for each setting, we have selected the best model in the first 20 epochs. Finally, the number of latent factors is equal to $20$. This value reflects a trade-off between latent factors' expressiveness and memory space limits (given by a realistic Federated Learning environment).
We have measured the recommendation accuracy by exploiting: Precision ($P@N$) (the proportion of relevant items in the recommendation list), and Recall ($R@N$), that measures the relevant suggested items.
Regarding diversity, we have adopted Item Coverage ($IC$) and Gini Index ($G$). The former provides the overall number of diverse recommended items, and it highlights the degree of personalization expressed by the model \cite{DBLP:journals/tkde/AdomaviciusK12}. The latter measures how unequally an RS provides users with different items \cite{DBLP:reference/sp/CastellsHV15}, being higher values corresponding to more tailored lists. 

\subsection{Performance of Federated Learning to Rank}
\begin{table*}[t]
\scriptsize
\caption{Recommendation performance for baselines and \framework on the three datasets. For each value of $T$, the experiment with the best $\pi$ is shown.}
\centerline{
{\setlength{\tabcolsep}{0.3em}
\begin{tabular}{llrrrrrrrrrrrr}
\toprule
 &  & \multicolumn{ 4}{c}{\textbf{Amazon Digital Music}} & \multicolumn{ 4}{c}{\textbf{LibraryThing}} & \multicolumn{ 4}{c}{\textbf{MovieLens 1M}} \\ 
   \cmidrule(r){3-6}
 \cmidrule(lr){7-10}
  \cmidrule(l){11-14}
 &  & \multicolumn{1}{l}{\textbf{P@10}} & \multicolumn{1}{l}{\textbf{R@10}} & \multicolumn{1}{l}{\textbf{IC@10}} & \multicolumn{1}{l}{\textbf{G@10}} & \multicolumn{1}{l}{\textbf{P@10}} & \multicolumn{1}{l}{\textbf{R@10}} & \multicolumn{1}{l}{\textbf{IC@10}} & \multicolumn{1}{l}{\textbf{G@10}} & \multicolumn{1}{l}{\textbf{P@10}} & \multicolumn{1}{l}{\textbf{R@10}} & \multicolumn{1}{l}{\textbf{IC@10}} & \multicolumn{1}{l}{\textbf{G@10}} \\ \midrule

\textbf{Random} &  & 0.00005 & 0.00005 & 14186 & 0.28069 & 0.00054 & 0.00028 & 31918 & 0.60964 & 0.00871 & 0.00283 & 3666 & 0.85426 \\ \hline
\textbf{Most Popular} &  & 0.00469 & 0.00603 & 24 & 0.00023 & 0.05013 & 0.03044 & 36 & 0.00031 & 0.10224 & 0.03924 & 118 & 0.00569 \\ \hline
\textbf{User-kNN} &  & 0.01940 & 0.02757 & 4809 & 0.04115 & 0.14193 & 0.10115 & 3833 & 0.01485 & 0.12613 & 0.06701 & 737 & 0.04636 \\ \hline
\textbf{Item-kNN} &  & 0.02147 & 0.03171 & 4516 & 0.03801 & 0.20214 & 0.14778 & 12737 & 0.09979 & 0.08873 & 0.05475 & 2134 & 0.19292 \\ \hline
\textbf{VAE} &  & 0.01580 & 0.02289 & 3919 & 0.04179 & 0.10834 & 0.07711 & 7800 & 0.04638 & 0.11735 & 0.06192 & 1476 & 0.09259 \\ \toprule
\textbf{BPR-MF} &  & 0.00921 & 0.01298 & 739 & 0.00415 & 0.07009 & 0.04303 & 3082 & 0.01359 & \textbf{0.11911} & 0.05817 & \textbf{1444} & \textbf{0.08508} \\ \hline
\textbf{FCF} &  & 0.00839 & 0.01222 & \textbf{2655} & 0.01861  & \textbf{0.10760} & 0.04392 & 829 & 0.01305  & 0.10760 & 0.04392 &  829 & 0.01305  \\ \hline
\multicolumn{ 1}{l}{\textbf{\framework}} \\ \cline{ 2- 14}
\multicolumn{ 1}{l}{\textbf{$T=1$ }} & & 0.00610 & 0.00889 & 349 & 0.00136 & 0.06309 & 0.03738 & 1650 & 0.00512 & 0.11805 & \textbf{0.05902} & 1041 & 0.06608 \\ \cline{ 2- 14}
$T=X^+/|\mathcal{U}|$  & & \textbf{0.01422} & \textbf{0.02060} & 2586 & \textbf{0.02153} & 0.08512 & \textbf{0.05627} & \textbf{5404} & \textbf{0.02784} & 0.11599 & 0.05571 & 1326 & 0.02513 \\ \bottomrule
\end{tabular}
}
}
\label{tab:general}
\end{table*}

We begin our experimental evaluation by investigating the efficacy of \framework, and we assess whether its performance is comparable to baseline algorithms. Table \ref{tab:general} depicts the results in terms of accuracy and diversity.
The Table is visually split into two parts. The algorithms in the bottom part (BPR-MF, FCF, and the two settings of \framework) are the factorization-based models. The upper part provides the positioning of \framework to the other state-of-the-art approaches.
Focusing on the factorization-based methods, we can note that BPR-MF outperforms \framework for $T=1$, but it remains at about 67\% and 88\% of the centralized algorithm for \textit{Amazon Digital Music} and \textit{LibraryThing}, respectively. 
However, the realistic Federated setting is with $T=X^+/|\mathcal{U}|$. Here, \framework consistently improves the recommendation performance with respect to BPR-MF and FCF, over the three datasets. Actually, for \textit{Amazon Digital Music} and \textit{LibraryThing} \framework improves accuracy metrics of about 50\% and 25\% with respect to BPR-MF. The achievement can be explained as an advantage brought by the increased local computation. It is worth noticing that these results partially contradict Rendle\textit{ et al.}~\cite{DBLP:conf/uai/RendleFGS09} since they hypothesize that traversing user-wise the training triples would worsen the recommendation performance. The same accuracy improvements are not visible in \textit{MovieLens 1M}, where we witness results comparable or worse than BPR-MF, probably due to the overfitting caused by the very high ratio between ratings and items.
\framework with increased computation still results robust with respect to the $IC$ metric, since, in general, it outperforms or remains comparable to FCF and BPR-MF.

\subsection{Analysis of Positive Feedback Transmission Ratio}
\label{sec:tuning}
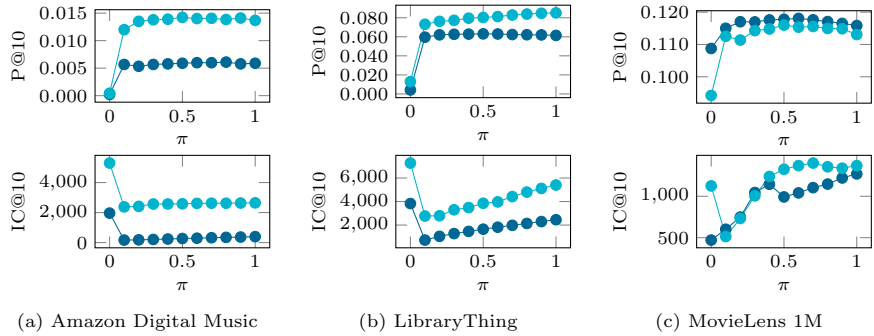
\begin{figure}[t]
\scriptsize
\captionsetup[sub]{labelsep=space,font=scriptsize}

\begin{subfigure}{.32\textwidth}
  \centering
    \begin{tikzpicture}
    \begin{axis}[
    width=1\linewidth,
    height=80pt,
    yticklabel style={
    /pgf/number format/.cd, fixed, fixed zerofill,
    /pgf/number format/precision=3
    },
    scaled y ticks=false,
    xlabel=$\pi$,
    ylabel=P@10,
    ylabel near ticks,
    xtick={0,0.5,1},
    xlabel near ticks
    ]
    \addplot [MidnightBlue, mark=*, restrict expr to domain={\coordindex}{0:10}] table [x=pi, y=ADM_P, col sep=tab] {csv/varying_pi.csv};
    \addplot [Turquoise, mark=*, restrict expr to domain={\coordindex}{11:21}] table [x=pi, y=ADM_P, col sep=tab] {csv/varying_pi.csv};
    \end{axis}
    \end{tikzpicture}
\end{subfigure}
\begin{subfigure}{.32\textwidth}
  \centering
    \begin{tikzpicture}
    \begin{axis}[
    width=1\linewidth,
    height=80pt,
    yticklabel style={
    /pgf/number format/.cd, fixed, fixed zerofill,
    /pgf/number format/precision=3
    },
    scaled y ticks=false,
        xlabel=$\pi$,
            ylabel=P@10,
    ylabel near ticks,
        xtick={0,0.5,1},
            xlabel near ticks
    ]
    \addplot [MidnightBlue, mark=*, restrict expr to domain={\coordindex}{0:10}] table [x=pi, y=LT_P, col sep=tab] {csv/varying_pi.csv};
    \addplot [Turquoise, mark=*, restrict expr to domain={\coordindex}{11:21}] table [x=pi, y=LT_P, col sep=tab] {csv/varying_pi.csv};
e    \end{axis}
    \end{tikzpicture}
\end{subfigure}
\begin{subfigure}{.32\textwidth}
  \centering
    \begin{tikzpicture}
    \begin{axis}[
    width=1\linewidth,
        height=80pt,
    yticklabel style={
    /pgf/number format/.cd, fixed, fixed zerofill,
    /pgf/number format/precision=3
    },
    scaled y ticks=false,
        xlabel=$\pi$,
            xtick={0,0.5,1},
            ylabel=P@10,
    ylabel near ticks,
        xlabel near ticks
    ]
    \addplot [MidnightBlue, mark=*, restrict expr to domain={\coordindex}{0:10}] table [x=pi, y=ML_P, col sep=tab] {csv/varying_pi.csv};
    \addplot [Turquoise, mark=*, restrict expr to domain={\coordindex}{11:21}] table [x=pi, y=ML_P, col sep=tab] {csv/varying_pi.csv};
    \end{axis}
    \end{tikzpicture}
\end{subfigure}

\begin{subfigure}{.32\textwidth}
  \centering
    \begin{tikzpicture}
    \begin{axis}[
    width=1\linewidth,
    height=80pt,
    yticklabel style={
    /pgf/number format/fixed,
    /pgf/number format/precision=2
    },
    scaled y ticks=false,
    xlabel=$\pi$,
    ylabel=IC@10,
    ylabel near ticks,
    xtick={0,0.5,1},
    xlabel near ticks
    ]
    \addplot [MidnightBlue, mark=*, restrict expr to domain={\coordindex}{0:10}] table [x=pi, y=ADM_IC, col sep=tab] {csv/varying_pi.csv};
    \addplot [Turquoise, mark=*, restrict expr to domain={\coordindex}{11:21}] table [x=pi, y=ADM_IC, col sep=tab] {csv/varying_pi.csv};
    \end{axis}
    \end{tikzpicture}
  \caption{Amazon Digital Music}
\end{subfigure}
\begin{subfigure}{.32\textwidth}
  \centering
    \begin{tikzpicture}
    \begin{axis}[
    width=1\linewidth,
    height=80pt,
    yticklabel style={
    /pgf/number format/fixed,
    /pgf/number format/precision=2
    },
    scaled y ticks=false,
        xlabel=$\pi$,
            ylabel=IC@10,
    ylabel near ticks,
        xtick={0,0.5,1},
            xlabel near ticks
    ]
    \addplot [MidnightBlue, mark=*, restrict expr to domain={\coordindex}{0:10}] table [x=pi, y=LT_IC, col sep=tab] {csv/varying_pi.csv};
    \addplot [Turquoise, mark=*, restrict expr to domain={\coordindex}{11:21}] table [x=pi, y=LT_IC, col sep=tab] {csv/varying_pi.csv};
    \end{axis}
    \end{tikzpicture}
  \caption{LibraryThing}
\end{subfigure}
\begin{subfigure}{.32\textwidth}
  \centering
    \begin{tikzpicture}
    \begin{axis}[
    width=1\linewidth,
        height=80pt,
    yticklabel style={
    /pgf/number format/fixed,
    /pgf/number format/precision=2
    },
    scaled y ticks=false,
        xlabel=$\pi$,
            xtick={0,0.5,1},
            ylabel=IC@10,
    ylabel near ticks,
        xlabel near ticks
    ]
    \addplot [MidnightBlue, mark=*, restrict expr to domain={\coordindex}{0:10}] table [x=pi, y=ML_IC, col sep=tab] {csv/varying_pi.csv};
    \addplot [Turquoise, mark=*, restrict expr to domain={\coordindex}{11:21}] table [x=pi, y=ML_IC, col sep=tab] {csv/varying_pi.csv};
    \end{axis}
    \end{tikzpicture}
  \caption{MovieLens 1M}
\end{subfigure}

\caption{F1 performance at different values of $\pi$ in the range $[0.1,1]$. Dark blue is $T=1$, light blue is $T=X^+/|\mathcal{U}|$.}
\vspace{-0.2cm}
\label{fig:varyingpi}
\end{figure}
We have extensively analyzed the behavior of \framework when tuning $\pi$ for sending progressive fractions of positive feedback in $[0.0, \dots, 1.0]$ with step $0.1$. 
We believe that the most important dimensions for this analysis are accuracy (Precision), and aggregate diversity (Item Coverage).  Figure~\ref{fig:varyingpi} reports the results for the two experimented settings.
Even here, \adm and \library show similar trends.
The accuracy of the recommendation progressively increases reaching the maximum with fractions $0.8$ and $0.5$, respectively, for $T=1$, and with fractions $0.9$ and $1.0$ for $T=X^+/|\mathcal{U}|$.
First, this suggests that, at the beginning of the training, some positive feedback is needed for establishing the value of an item.
Notwithstanding, even with $\pi=0.1$ (i.e., sharing just 10\% of private information), we witness a jump in recommendation accuracy (one order of magnitude), reaching up to 92\% of the best accuracy.
We should also observe another significant behavior. 
With a fraction of $0.0$, we observe a high value of \textit{IC}, with poor recommendation accuracy.
It suggests that the system could not capture population preferences, and it behaves similarly to Random.
However, even with a small fraction of positive feedback like $0.1$, we observe a significant decrease in diversity metrics.
The system learns which items are popular and starts suggesting them.
Moreover, if we observe large fractions, we may notice that diversity increases as we feed the system with more information.
For \ml, it is worth noticing that \framework shows accuracy performance extremely close to the best value by sharing only $10\%$ of positive interactions.
This behavior may be due to several reasons.
Firstly, \ml is a relatively dense dataset in the recommendation scenario (it has a sparsity of $0.955$).
Secondly, it shows a very high user-item ratio \cite{DBLP:journals/tmis/AdomaviciusZ12} (i.e., $1.63$) compared to \adm ($0.04$), and \library ($0.20$), and it shows high values for the average number of ratings per user ($132.87$), and ratings per item ($216,56$).
All these clues suggest that the system learns how to rank items even without the need for the totality of ratings.
If we focus on diversity metrics, $IC$ and Gini, we may notice that diversity is progressively increasing from fraction $0.1$ to $1.0$.
It suggests that the system recommends a small number of popular items with a fraction of $0.1$, while it provides more diversified recommendation lists considering larger portions of positive user feedback.
At  this  stage  of  the  analysis, we can draw an interesting consideration: in general, the highest accuracy values do not correspond to the fraction of $1.0$.
Specifically, the experiments show that, initially, the recommender struggles to suggest relevant items without positive feedback (fraction $0.0$).
However, with a small injection of feedback, the system starts to work well.
Nonetheless, in \adm and \library, if we increase the fraction, we witness an increase concerning accuracy only until a certain value of $\pi$.
Although this consideration, we observe an increase in diversity metrics when we continue to increase the value of $\pi$.
Since it has a small or even detrimental impact on accuracy, those items might be unpopular items erroneously suggested to users.

\begin{figure*}[!t]
    \centering
    \centerline{\includegraphics[width=1.05\linewidth]{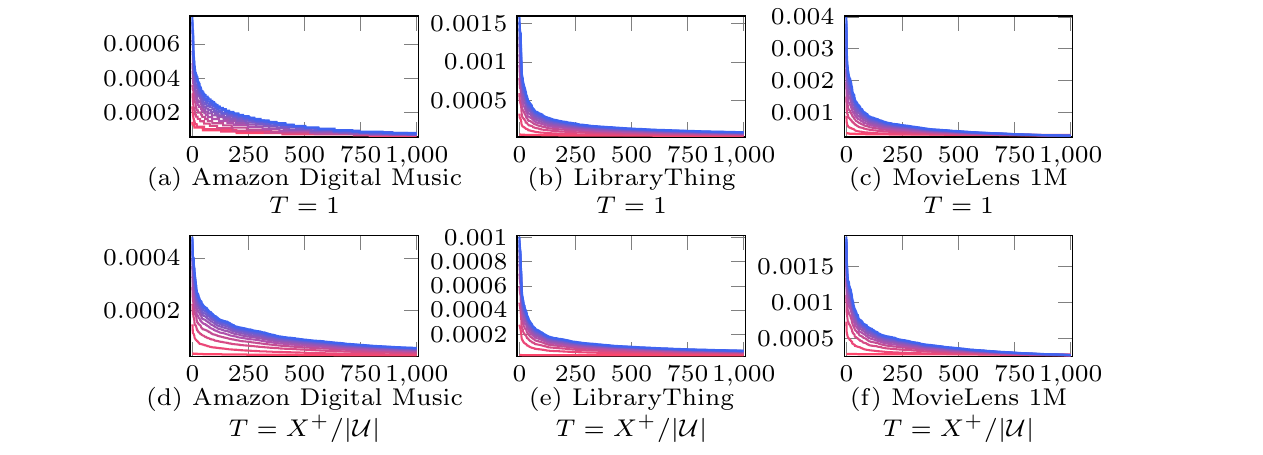}}
    \caption{Normalized number of item updates during the training: the 1,000 most updated items for different values of $\pi$ (from $\pi=0.0$ in red to $\pi=1.0$ in blue).}\label{fig:exchanged}
\end{figure*}

\begin{figure*}[!t]
    \centering
    \centerline{\includegraphics[width=1.05\linewidth]{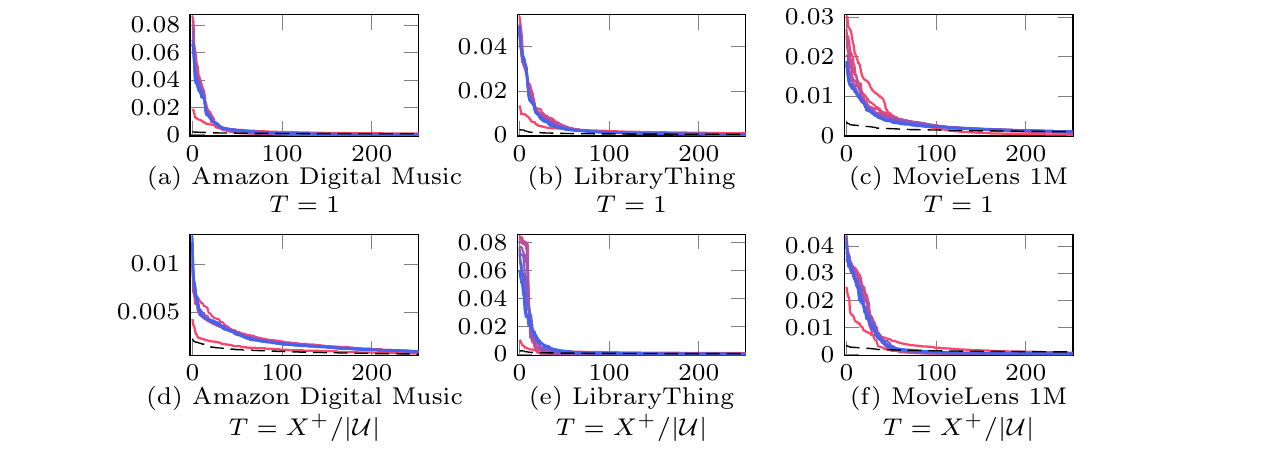}}
    \caption{Normalized number of recommendations for each item (colored curves from $\pi=0.0$ in red to $\pi=1.0$ in blue) vs. normalized amount of positive feedback per item (black dashed curve). The 250 most popular items are shown.}\label{fig:recommended}
\end{figure*}

\subsection{Study on \framework algorithmic bias}\vspace{-0.5em}
In this section, we study how incomplete transmission of user feedback affects the item popularity in the recommendations and during the learning process.
It is essential to discover whether the exploitation of a FL approach influences the algorithmic bias, determining popular items to be over-represented~\cite{DBLP:conf/recsys/Baeza-Yates20,DBLP:conf/sigir/CanamaresC18}.
We have re-trained \framework with all the previously considered $\pi$.
For each experiment, we analyzed the data flow between the clients and the server.
Afterward, we have extracted the number of updates for each item. 
Figure~\ref{fig:exchanged} illustrates the occurrences for the $1,000$ most updated items.
In the Figure, the curve colors denote the different $\pi$, while the values represent the update frequency during the training process for each item on the horizontal axis. Analogously, we considered the final top-10 recommendation list of each user.
Following the same strategy, we analyzed the occurrences of the items in the recommendation. 
Then, we ordered items from the most to the least recommended, and we plotted the occurrences of the first 250 in Figure~\ref{fig:recommended}.
To compare the different datasets, we have normalized the values considering the overall dataset occurrences.
Figure~\ref{fig:exchanged} shows that data disclosure, i.e., the value of $\pi$, highly influences the information exchanged during the training process.
Additionally, the update frequency curve exhibits a constant behavior for all the datasets, when $\pi = 0.0$. 
This trend suggests that items are randomly updated without taking into account any information about item popularity. This behavior explains the high \textit{IC} entirely observed in Figure~\ref{fig:varyingpi} for $\pi = 0.0$.
The curve for $\pi=0.1$ shows that the exchanged data is enough to provide the system with information about item popularity.
The curves suggest that the information on item popularity is being injected into the system. 
By increasing the value of $\pi$, the trend becomes more evident.
Due to the original rating distribution, the system initially exchanges more information about the very popular items.
To analyze the algorithmic bias, we can observe Figure~\ref{fig:recommended}, where the colored curves represent the frequency of item recommendation on the horizontal axis, and the black dashed curve the amount of positive feedback for that item in the dataset. 
Remarkably, item popularity in recommendation lists does not vary as we may expect based on the previous analysis.
The setting $\pi=0.0$ is an exception, as extensively explained before.
Since in \adm and \library the updates sent by the clients are randomly selected among the negative items, \framework acts like a Random recommender.
The system cannot catch popularity information and
it struggles to make the right items popular. 
Finally, we can focus on the curves for  $\pi > 0$. 
It is noteworthy that the $\pi$ curves behave similarly, and they propose the same proportion of popular items. 
The curves show the model absorbs the initial variation in exchanged item distribution, unveiling an unknown aspect of factorization models.

\section{Conclusion and Future Work}
\label{sec:conclusions}
In this paper, we have tackled the problem of putting users in control of their private data for a recommendation scenario. Witnessing the growing concern about privacy, users might want to exploit their sensitive data and share only a small fraction of it.
In such a context, classic CF approaches are no more feasible. To overcome these problems, we have proposed \framework, a novel recommendation framework that respects the FL paradigm. 
With \framework, private user feedback remains on user devices unless they decide to share it. Nevertheless, \framework ensures high-quality recommendations despite the constrained setting.
We have extensively studied the performance of \framework by comparing it with other state-of-the-art methods.
We have then analyzed the impact of progressive reduction of user feedback and studied the effects on the diversity of the recommendation results. 
Finally, we have investigated whether the federated algorithm imposes an algorithmic bias to the recommended lists. 
The study paves the way for further research directions. 
On the one hand, the results' analysis suggests that centralized recommender systems are not performing at their best. Feeding recommender systems with all the available feedback, without any filtering, may lead to a performance worsening. On the other hand, the competitive results of \framework suggest that the FL-based algorithms show a recommendation quality that makes them suitable to be adopted on a massive scale.

\bibliographystyle{abbrv}
\bibliography{bibliography}
\end{document}